\documentclass[aps,showpacs,twocolumn,floats,superscriptaddress]{revtex4}
\usepackage{graphicx}
\usepackage{dcolumn}
\usepackage{bm}
\usepackage{color}
\usepackage{amssymb}

\begin{document}
\author{M. Guglielmino}
\affiliation{Dipartimento di Fisica, Politecnico di Torino, Corso Duca degli Abruzzi 24, I-10129 Torino, Italy}
\affiliation{Institute for Theoretical Atomic, Molecular and Optical Physics,
Harvard-Smithsonian Center of Astrophysics, Cambridge, MA, 02138}
\author{V. Penna}
\affiliation{Dipartimento di Fisica and CNISM Unit\`a di Ricerca, Politecnico di Torino, Corso Duca degli Abruzzi 24, I-10129 Torino, Italy}
\author{B. Capogrosso-Sansone}
\affiliation{Institute for Theoretical Atomic, Molecular and Optical Physics,
Harvard-Smithsonian Center of Astrophysics, Cambridge, MA, 02138}

\title{Ising antiferromagnet with ultracold bosonic mixtures confined in a harmonic trap}

\begin{abstract}
We present accurate results based on Quantum Monte Carlo simulations of two-component bosonic systems on a square lattice and in the presence of an external harmonic confinement. Starting from hopping parameters and interaction strengths which stabilize the Ising antiferromagnetic phase in the homogeneous case and at half integer filling factor, we study how the presence of the harmonic confinement challenge the realization of such phase. We consider realistic trapping frequencies and number of particles, and establish under which conditions, i.e. total number of particles and population imbalance, the antiferromagnetic phase can be observed in the trap. 
\end{abstract}


\maketitle

\begin{figure}%
\begin{center}
\includegraphics[width=\columnwidth]{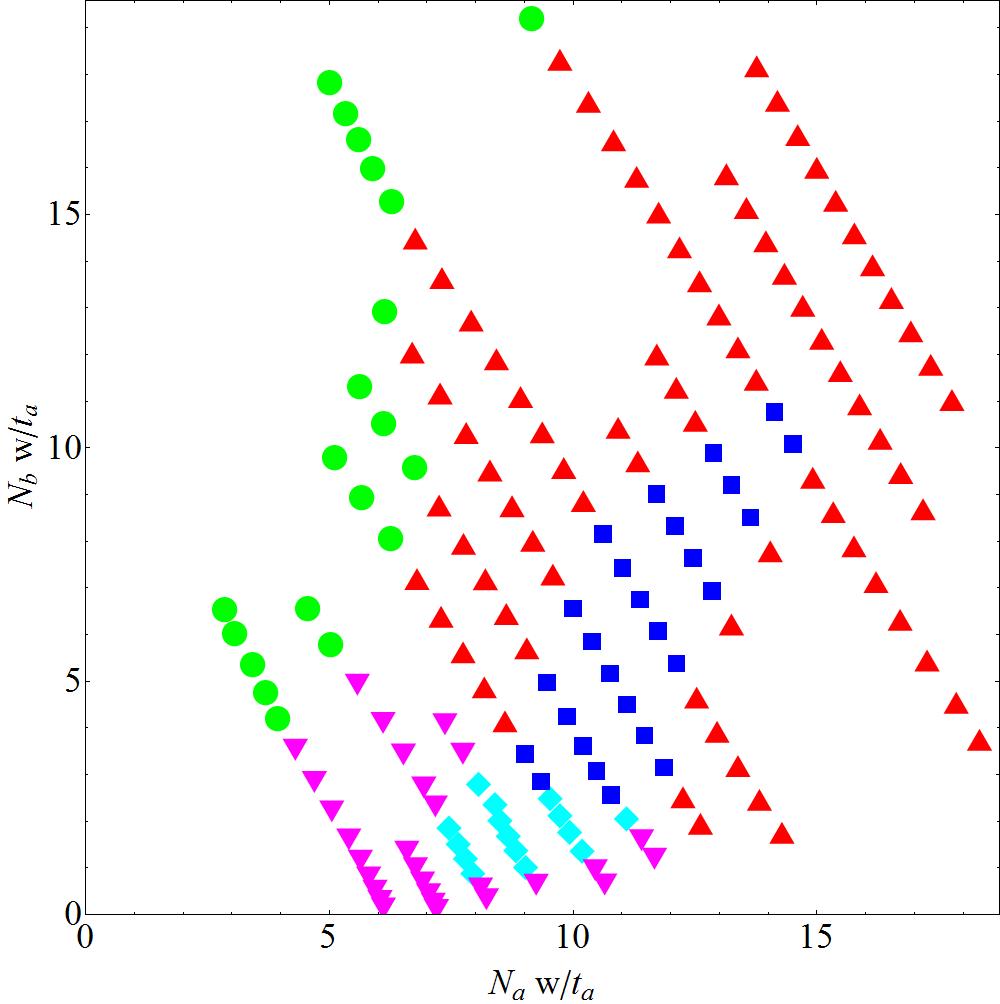}%
\end{center}
\caption{(Color online) Typologies of phase coexistence (see text) as a function of  number of particles 
$N_a$, $N_b$ scaled by the trap curvature for $w=0.007, \; 0.005, \; 0.0025\; t_a$. Diamonds: 
CB core; squares: CB ring; downward triangles : global Mott core; upward triangles: 
global Mott ring; circles: phase separation. Error bars are much smaller than symbol size 
(maximum percent error is 0.5\%). }
\label{fig:phase_diagram}%
\end{figure}

In recent years great attention has been devoted to mixtures of ultracold atomic gases in
optical lattices~\cite{Ospelkaus, Esslinger, Bloch, Bloch2, Minardi1, Schneble, Weld1, Weld2}. 
One of the most remarkable features of such systems is the possibility of stabilizing quantum 
magnetic phases~\cite{Demler_Lukin,Kuklov_Svistunov}. Though interesting and fundamental on
their own, better understanding of magnetic phases is further motivated by applications to
quantum-information processing and their relevance to unconventional superconductivity such as
high-temperature and heavy-fermion superconductivity. It has been proposed~\cite{AFandSC} (and
recently experimentally investigated~\cite{Si}) that antiferromagnetic excitations may be responsible
for the pairing mechanism in unconventional  superconductors. The experimental realization of magnetic Hamiltonians with ultracold atoms would open up the way for direct control over interactions, geometry,
and frustration. Recently an Ising antiferromagnetic phase has been realized with single-component
ultracold atoms trapped in a one-dimensional optical lattice tilted by 
a magnetic field~\cite{tilted_lattice} (Ising density wave order could
also be realized at lower energy scales in specific two-dimensional
tilted geometries~\cite{Pielawa}). 
\\ \indent Magnetic phases with mixtures of ultracold atoms can be realized at low temperatures,
strong interactions and specific filling factors. Experimentally, though, the presence of the
external confinement will challenge their realization. Previous theoretical studies, focused on
Bose-Bose mixtures, reported accurate results for the ground-state phase boundaries of the Ising antiferromagnetic and the \emph{xy}-ferromagnetic  phases at half-filling factor~\cite{Soyler},
and for the critical temperature needed for their realization~\cite{Capogrosso_Soyler}.
The temperature scale, determined by spin-exchange interactions, results challengingly low
($\sim$~few hundreds pK). Nonetheless, recent experimental efforts have been devoted towards the
development of new techniques of refrigeration and thermometry~\cite{Weld1,Weld2}. 
\\ \indent
The next crucial question to theoretically address is studying the effect of an external trapping
potential. In the present work we focus on the Ising antiferromagnetic (AF) phase realized by Bose-Bose
mixtures in the strongly interacting regime and at half-filling factor, and study under which conditions
it survives the presence of a harmonic confinement. In the local-density approximation, the latter provides
a scan over chemical potentials of the two components. Therefore, even if Hamiltonian parameters are experimentally tuned to stabilize the Ising AF phase in the homogeneous system, the presence of an external confinement will challenge its realization. 
In other words, other phases such as Mott insulator (MI) + superfluid, double superfluid and phase separation will compete, moreover given the pronounced asymmetry in hopping parameters required for stabilization of the AF 
phase~\cite{Demler_Lukin,Kuklov_Svistunov,Soyler}.
%
%
%
Hence, for a given harmonic confinement, the challenge
consists in tuning the particle numbers 
%
%
%
%
%
%
so that there exist an extended region in the trap with half filling
for both components.
\begin{figure*}
\begin{center}
\includegraphics[width=\textwidth]{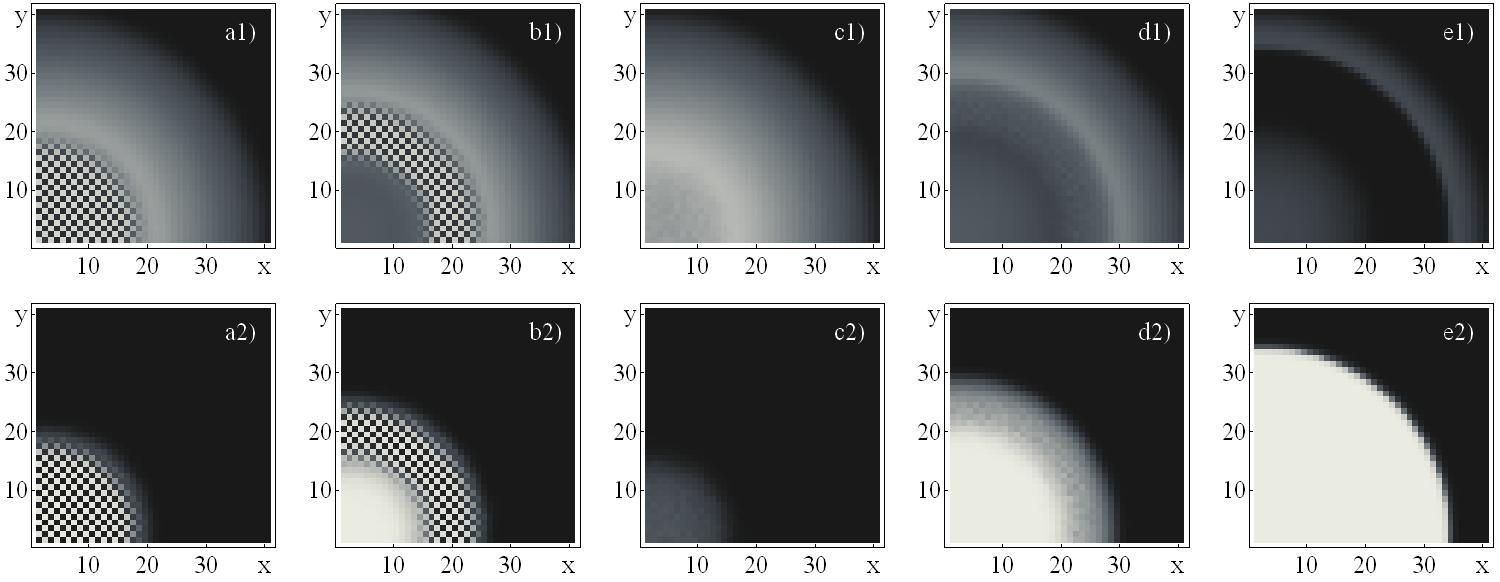} \\%
\includegraphics[width=\textwidth]{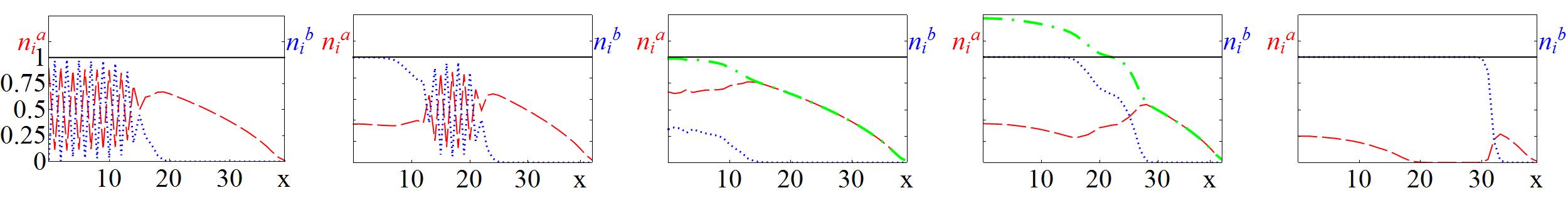}
\end{center}
\caption{(Color online) Top and middle rows: representative density maps (light and heavy component respectively)
for each
type of phase coexistence observed. Color scheme: white (black) corresponds to
1 (0) density. Bottom row: density profiles along a line going through the trap center (dashed red, 
dotted blue
and dot-dashed green lines correspond to light component, heavy component and total density profiles,
respectively). Coordinates $x$ and $y$ are given in units of $d$.  Hamiltonian parameters:  $t_b=0.15t_a$, $U_{ab}=5.7t_a$, $T=0.05t_a$, 
$w=0.005 t_a$. First column: 2CB core, $N_a=1940$, $N_b=430$; second column: 2CB ring,
$N_a=2155$, $N_b=1030$; third column: global Mott core, $N_a=2130$, $N_b=125$; 
fourth column: global Mott ring, $N_a=1750$, $N_b=1750$; fifth column:
phase separation, $N_a=425$, $N_b=3185$.}
\label{fig:dens_map}
\end{figure*}
\\ \indent
In the following we present first precise results based on Quantum Monte Carlo simulations of
a two-dimensional system with numbers of particles typical of current experiments~\cite{Greiner,Bloch3}
and trapping frequencies easily realizable with magnetic or optical traps. We show that  observation of the
AF phase in a trap is non trivial, but indeed possible, provided one properly tunes the total number of
particles and the population imbalance. Hence, our results provide an accurate guidance towards the
experimental realization of the Ising AF with mixtures. 
\\ \indent
We consider a system of two-component bosons in a square lattice,  with repulsive interspecies
interaction and in the presence of an external harmonic confinement. This system can be realized
by loading an optical lattice with two different atomic species~\cite{Minardi1}, or the same atomic
species in two different internal energy states~\cite{Schneble,Weld1,Weld2}. If intraspecies
interactions are much larger than any other energy scale (a limit which can be achieved by, e. g.,
using Feshbach resonance%
), and the temperature is low enough, the system is accurately
described by the two-component hard-core Bose-Hubbard Hamiltonian:
$$
H = - \sum_{\langle i j \rangle} {\left ( t_a \, a_i^\dag a_j + t_b \, b_i^\dag b_j \right )} -         
\sum_{\alpha ,\, i} { \mu^\alpha_i \, n_i^{\alpha}} \nonumber  \\
%
%
+ U_{ab} \sum_i { n_i^{a} \! n_i^{b} }, 
$$
where $\alpha\!=\!a,b\,$ denotes the two species; 
${\sum}_{\langle i j \rangle}$ is intended on nearest-neighboring sites; 
$a_i$ ($a_i^\dag$),  $b_i$ ($b_i^\dag$) are bosonic annihilation (creation)
operators at site $i$, satisfying the hard-core constraint; $t_a$, $t_b$ are hopping amplitudes; 
$n_i^{a}\!=a_i^\dag a_i$, $n_i^{b}\!=b_i^\dag b_i,$ are number operators;
$U_{ab}$ is the interspecies interaction; 
$\mu^\alpha_i=\mu_\alpha - w ( {\bf r}_i/d )^2$, with $\mu_\alpha$ the chemical potential,
$\bf{r}_i$ the vector position of site $i$, $d$ the lattice spacing, 
and $w$ the curvature of the external potential. The usual intraspecies interaction term 
$H_{int}={\sum}_{\alpha \, i} { U_\alpha \, n_i^{\!\alpha} \! ( n_i^{\alpha} - 1 )/2 } $
has been dropped because the $U_\alpha \to \infty$ limit is considered.
In the following we use  $t_a$ as the energy unit. 
\\ \indent In the strongly-interacting limit, Hamiltonian $H$ can be mapped 
onto an anisotropic Heisenberg Hamiltonian thus realizing quantum magnetic 
phases~\cite{Demler_Lukin,Kuklov_Svistunov}. In the following we focus on the Ising AF phase
which, in bosonic language, is equivalent to a double checker-board (2CB) solid, characterized
by broken $Z_2$ symmetry. To this end, we consider the hard-core boson regime set $t_b=0.15t_a$, $U_{ab}=5.7t_a$, so that the
ground state of the homogenous system at half-integer filling corresponds to the 2CB
phase~\cite{Soyler}. We notice that the 2CB phase can be realized only for
strong enough asymmetry between $t_a$ and $t_b$, i.e. if component 
B is, via e.g. optical depths tuning, effectively much heavier than A.
\\ \indent We have performed Quantum Monte Carlo simulations by means of the Worm
Algorithm~\cite{Worm} of model $H$ for several curvatures of the harmonic potential $w/t_a=0.1$, 
$0.02$, $0.014$, $0.007$, $0.005$, $0.0025$, with maximum number of particles (corresponding
to the most shallow trap) $N=N_a+N_b\sim 7200$. For each $w$ and $N$ the system size was
chosen large enough so that no finite-size effects were present (ranging from linear size 
$L=60$ to $L=135$, depending on the specific case),  and the temperature was
fixed to $T=0.05t_a$, well below the 2CB phase-transition temperature for the chosen
Hamiltonian parameters~\cite{Capogrosso_Soyler}. Our goal is to establish for which
total particle number $N$ and population imbalance $(N_a-N_b)$ the 2CB phase is stabilized
in the trap. To this extent we systematically analyze and classify different typologies of
phase coexistence.
\\  \indent We have observed five distinct types of phase coexistence, as illustrated in
Fig.~\ref{fig:dens_map}, where the top (middle) row refers to the in trap density maps of the
light (heavy) component, and the bottom row shows corresponding density profiles along a line
going through the center of the trap. The color scheme in the top and middle rows is such that
white (black) corresponds to density 1 (0). In the bottom row, dashed red, dotted blue and dot-dashed green refer to
light component, heavy component and total density profiles, respectively. 
Each column is representative of an extended region in the plot of Fig.~\ref{fig:phase_diagram}
summarizing the different types of phase coexistence found, as we shall discuss below. 
We classify the different typologies as follows (for each typology phases are listed as they
appear going from the center to the edges of the trap): 
\\ \indent \emph{2CB core}: 2CB, light component superfluid (Fig.~\ref{fig:dens_map}, panel a1-a2). 
\\ \indent \emph{2CB ring}: heavy component MI + light component superfluid, 2CB, light component
superfluid (Fig.~\ref{fig:dens_map}, panel b1-b2).
\\ \indent \emph{global Mott core}: global MI, i.e. a MI in the total density (see profile of
total density, green line in Fig.~\ref{fig:dens_map}, panel c3 ), light component superfluid
(see Fig.~\ref{fig:dens_map}, panel c1-c2).
\\ \indent \emph{global Mott ring}: heavy component MI + light component superfluid, global MI
(see profile of total density, green line in Fig.~\ref{fig:dens_map}, panel d3), 
light component superfluid (see Fig.~\ref{fig:dens_map}, panel d1-d2).
\\ \indent \emph{phase separation}: heavy component MI + light component superfluid, 
heavy component MI, light component SF (see Fig.~\ref{fig:dens_map}, panel e1-e2).
\\ For each of such sequences, subsequent phases are spatially separated by a thin shell
of double superfluid. On the bottom row, density profiles make clear the choice of classification
of typologies. In particular, in correspondence of the global MI phase (columns c) and d) ),
we have observed corrugations in the density profiles of the single components, which
have not completely disappeared within simulation time. 
We have observed that the super-counter-fluid phase~\cite{Kuklov_Svistunov} exhibits a
similar behavior, and suspect that the global MI phase could display an analogous
off-diagonal-long-range-order.
%
\\ \indent
In all our simulations we have observed that the heavy component tends to locate itself
at the center of the trap. This behavior can be understood as follows. In the strong-coupling
regime considered here, delocalization of the heavy component (at the expense of the energy
required to climb the walls of the trap) is suppressed compared with the delocalization of
the light component, due to larger hopping parameter of the latter. 
\\ \indent
Our results for the three shallower trapping potentials simulated
($w=0.007,\; 0.005,\; 0.0025\; t_a$) are summarized in Fig.~\ref{fig:phase_diagram}, 
where each symbol corresponds to one of the typologies described above (diamond for
2CB core, square for 2CB ring, downward triangle for global Mott core, upward triangle
for global Mott ring, circle for phase separation). 
We can identify $(N_a, N_b)$ values (we scale the number of particles by the curvature of the
trapping potential, see below) for which the 2CB phase  is realized in certain regions of the
trap: squares and diamonds, corresponding to 2CB ring and 2CB core, respectively. We note that,
in general, the 2CB phase can be observed only if the light component is the majority one, and
provided $w< 0.01\; t_a$. 
We have studied larger values of $w$ (= 0.014, 0.02, 0.1 $t_a$), i.e. steeper traps,  and have
not observed CB for any $(N_a, N_b)$. In other words, the trapping potential should be smooth
enough to allow stabilization of 2CB ring or core regions. For $w=0.007,\; 0.005,\; 0.0025\; t_a$,
instead, we can summarize our results on the same plot by using $w N_a /t_a$ and $w N_b/t_a$ as
axis (the smoother the trap, the wider 2CB core and ring regions are). Upon further decreasing
the trap curvature $w$ we have observed that 2CB ring region in the $w N_a/t_a$ vs $w N_b/t_a$
plane tends to grow, although extremely slowly (e.g. lowering $w$ by a factor of 2, would only
convert a single upward triangle into a square, i. e. a global MI ring would become a CB ring,
in the plot of Fig.~\ref{fig:phase_diagram}).  In other words, for \emph{much} shallower
trap curvatures, the CB ring region would be only slightly larger than the one in the plot
of Fig.~\ref{fig:phase_diagram} (i. e. the system exhibits  a scalable behavior for a wide range of $w$'s). 
Hence, Fig.~\ref{fig:phase_diagram} can be used as guidance for experiments since it provides
$N_a$ and $N_b$ required for the observation of the Ising AF for experimentally realizable trap
curvatures.
\\ \indent 
Although the hard-core limit corresponds to most stable quantum magnetic phases against 
hopping~\cite{Demler_Lukin}, we will now relax this constraint. We have performed simulations 
at finite intraspecies interaction $U_a=U_b=U$ both in uniform and trapped system. Our goal is
to probe how far one can soften $U$ for given values of hopping parameters $t_a$, $t_b$, and
interspecies interaction $U_{ab}$. We have found that for $t_b=0.15t_a$ and $U_{ab}=5.7t_a$ the
2CB phase in a uniform system survives down to $U\sim70t_a$, i.e. $U/U_{ab}\simeq12$. 
If we further increase the hopping asymmetry, e.g. $t_b=0.075t_a$, a 2CB phase still exists at
$U\sim40t_a$, i.e. $U/U_{ab}\simeq7$. We expect that the extension and position of 2CB regions
in Fig.\ref{fig:phase_diagram} will slightly change upon relaxing the hard-core constraint.
In particular, we expect the 2CB regions to shrink (due to a more fragile CB phase for soft-core
atoms) as phase separation becomes more favorable. 
%
%
%
Preliminary numerical results agree with such
an expectation but also suggest that the general qualitative results (i.e. typologies observed)
valid for hard-core bosons should still be valid. 
More specifically, we have performed simulations
in a trap with parameters $w/t_a=0.005$, $t_b=0.075t_a$, $U_{ab}=5.7t_a$, $U=40t_a$, $N_a=1240$
and $N_b=170$ and found that the 2CB core still exist. 
\\ \indent A previous mean-field study carried on for soft-core bosonic atoms claimed
realization of 2CB ring and core regions in a trap~\cite{Shrestha}. The parameters used
in~\cite{Shrestha} are the following: $w/t_a=0.064$, $t_b=0.16t_a$, $U_{ab}=5.7t_a$, 
$U_a=1.9t_a$, $U_b=37.5t_b$. We note that component A is rather soft. Preliminary numerical
simulations with the same parameters indicate that phase separation will occur both, with
and without the harmonic trapping potential (in the presence of the latter complete phase
separation, i.e. no overlap between the two components anywhere in the trap, seems to occur). 
Moreover, our numerical results for soft-core atoms in uniform systems, indicate that the
2CB phase will be destroyed upon softening  $U<70t_a$ with $t_b=0.15 t_a$. Hence, the
results of~\cite{Shrestha} seem to be rather controversial.
\\ \indent 
Finally, we would like to provide estimates for experimental setups required for the observation
of the Ising AF (we use a harmonic approximation around the minima of the optical lattice 
potential~\cite{Gerbier}). 
The hard-core limit can be achieved 
by realizing ${\rm a}_\alpha \gg {\rm a}_{ab}$ ($\alpha =a,b$), with  ${\rm a}_{\alpha}$ and
${\rm a}_{a b}$ intra- and interspecies scattering lengths, respectively. The latter inequalities
can be satisfied by e.g. making use of a combination of dc and rf magnetic fields~\cite{Tscherbul},
or dc electric fields~\cite{Krems} in order to independently tune two scattering lengths 
via Feshbach resonances. Alternatively, for what concerns $U_{ab}$ tuning, Wannier-function overlap 
in the presence of state-dependent lattices can be manipulated.
Harmonic frequencies for the trapping potential are given by $\omega^2_{\alpha}=2w/(m_{\alpha}a^2)$.
Let us consider laser beams with $\lambda=1064$nm and a mixture of $^{87}$Rb atoms in
hyperfine states $|1,-1\rangle$ (A component) and $|2,-2\rangle$ (B component), for which
${\rm a}_{a b}=98.09a_0$ ($a_0$ is the Bohr radius), ${\rm a}_{a}=100.4a_0$ and
${\rm a}_{b}=98.98a_0$. According to the choice in~\cite{choice1}, with a
harmonic confinement of $\omega_a=\omega_b\sim2\pi\; 370 $Hz (corresponding to $w=0.005 t_a$)
one can observe the 2CB core (ring) by loading $N_a\sim1700$ (2200), $N_b\sim300$ (800).
With $\omega_a=\omega_b\sim2\pi\; 250 $Hz (corresponding to $w=0.0025 t_a$), instead,
one can observe the 2CB core (ring) by loading $N_a\sim3400$ (4400), $N_b\sim600$ (1600). 
For shallower traps, e.g. $\omega \sim 2\pi\; 80$Hz, one should load $N_a\sim34000$ (44000)
and $N_b\sim6000$ (16000).
\\ \indent
In the case of $^{87}$Rb-$^{41}$K mixtures (we choose $^{87}$Rb as the effectively-light component
i. e. component A) one has ${\rm a}_{\mathrm{Rb}\mathrm{K}}=163a_0$, ${\rm a}_{\mathrm{Rb}}=99a_0$
and ${\rm a}_{\mathrm{K}}=65a_0$. According to the choice in~\cite{choice2}, with a harmonic confinement
of  $\omega_{\mathrm{Rb}}\sim2\pi\; 366$Hz and $\omega_{\mathrm{K}}\sim2\pi\; 525$Hz (corresponding to
$w=0.005 t_a$), the 2CB core (ring) is realized by loading $N_{\mathrm{Rb}}\sim1700$ (2200), $N_{\mathrm{K}}\sim300$ (800).
With a choice of $\omega_{\mathrm{Rb}}\sim2\pi\; 254$Hz and $\omega_{\mathrm{K}}\sim2\pi\; 381$Hz 
(corresponding to $w=0.0025 t_a$), the 2CB core (ring) is realized by loading 
$N_{\mathrm{Rb}}\sim3400$ (4400), $N_{\mathrm{K}}\sim600$ (1600).
\\ \indent 
In conclusion, we have presented accurate results based on Quantum Monte Carlo simulations of
two-component bosonic systems on a square lattice and in the presence of an external harmonic
confinement. Starting from hopping parameters and interaction strengths which stabilize the Ising antiferromagnetic in the homogeneous case and at half integer filling factor, we have studied how
the presence of the harmonic confinement challenges the realization of such phase. We have considered
realistic trapping frequencies and number of particles, and established under which conditions, i.e.
total number of particles and population imbalance, the antiferromagnetic phase can be observed in
the trap. With recently developed single atom imaging techniques, direct observation of the Ising
ferromagnetic phase is possible~\cite{Greiner, Bloch3}. Alternatively one can use Bragg spectroscopy
in order to probe density-density correlations~\cite{Fort}.  Given the challengingly low-temperature
scale associated to quantum magnetic phases, it would be certainly interesting to study signatures
of the onset quantum magnetic phases on the approach of the critical temperature from above.
\\ \\ \indent 
We would like to thank F. Minardi, S. S\"oyler and S. Kuhr for fruitful discussions.
This work was supported by the Institute for Atomic, Molecular and Optical Physics (ITAMP). 
MG wants to thank ITAMP for warm hospitality and financial support. 


\begin{thebibliography}{50}
\bibitem{choice1} 
If lattice depths (in units of recoil energy) are $s_a\sim16$, $s_b\sim27$, one obtains $t_b\simeq0.15t_a$, $U_{aa}\simeq70t_a$, $U_{bb}\simeq100t_a$ ($U_{bb}\simeq667t_b$), $t_a\simeq11.6$Hz. With a shift between state dependent lattices of $\delta x/d=0.345$ one has $U_{ab}\simeq 5.7t_a$.
\bibitem{choice2} 
If $s_{\mathrm{Rb}}\sim16$, $s_{\mathrm{K}}\sim31$ (in units of the respective recoil energies) and scattering length $a_s^{(\mathrm{Rb}\,\mathrm{K})}$ is tuned via Feshbach resonance~\cite{Minardi2} to $4.1a_0$ one obtains $t_b\simeq0.15t_a$, $U_{ab}\simeq5.7t_a$, $U_{aa}\simeq70t_a$, $U_{bb}\simeq1064t_b$ ($U_{bb}\simeq160t_a$), $t_a\simeq11.5$Hz. Alternatively, instead of tuning $a_s^{(\mathrm{Rb}\,\mathrm{K})}$ one could shift the two lattices by $\delta x/a=0.40$.
\end{thebibliography}
\end{document}